\documentclass{article}
\usepackage[a4paper, scale = 0.7, centering]{geometry}
\usepackage{amssymb,amsmath,amsthm}
\usepackage{mathtools}
\usepackage{graphicx,epstopdf}
\usepackage[usenames]{color}
\usepackage{url,hyperref}
\usepackage{algorithm,algorithmic}
\usepackage{verbatim}
\usepackage[utf8]{inputenc}
\usepackage{siunitx}
\usepackage{graphicx}
\usepackage{subfig}
\usepackage{flafter}
\usepackage{cite}
\usepackage{float}
\usepackage{todonotes}

\newtheorem{remark}{Remark}[section]

\numberwithin{equation}{section}

\newcommand{\matlab}{\textsc{Matlab}}
\newcommand{\kwave}{\textsc{k-Wave}}

\newcommand{\fenics}{\textsc{FEniCS}}

\title{Feasibility of Acousto-Electric Tomography}

\author{Bjørn Jensen\footnote{bjorn.c.s.jensen@jyu.fi; \url{https://orcid.org/0000-0002-4743-2631}}  , Adrian Kirkeby\footnote{adrian@simula.no; \url{https://orcid.org/0000-0003-2741-7423}} , and Kim Knudsen\footnote{kiknu@dtu.dk; \url{https://orcid.org/0000-0002-4875-3074}}
\\ 
\vspace{1mm}\\
Faculty of Information Technology$^{*}$\\
University of Jyväskylä\\
Finland\\ \\ \\
Department of Numerical Analysis and Scientific Computing$^{\dagger}$\\ Simula Research Laboratory\\Oslo, Norway \\ \\ \\ Department of Applied Mathematics and Computer Science$^{* \dagger \ddagger}$\\
Technical University of Denmark\\
Kgs.\ Lyngby, Denmark }

\begin{document}

\date{}
\maketitle

\begin{abstract}
In acousto-electric tomography the goal is to reconstruct the electric conductivity in a domain from electrostatic boundary measurements of corresponding currents and voltages, while the domain is penetrated by a time-dependent acoustic wave.  In this paper we provide an in-silico model of the entire coupled physics scenario and perform computational experiments. We propose a complete inversion framework for acousto-electric tomography in two steps: First the interior power density is obtained from boundary measurements by solving a linear inverse problem; second the interior conductivity is reconstructed from the power density by solving a non-linear, fairly well-posed problem. We develop a numerical model with realistically chosen parameters inspired by medical imaging. The critical signal strength is analyzed and the omnipresent Johnson-Nyquist noise is estimated. With this setup we perform numerical experiments on synthetic data with noise and demonstrate the feasibility of the method. However, our findings are a mix of positive and negative. We reconstruct features even under severe noise conditions, but we also find that the required Signal-to-Noise ratio yet remains infeasible for practical purposes.
\end{abstract}

\newpage
\section{Introduction}
In a variety of applications in imaging science it is important to reliably image the electrical conductivity in some object or region. In medical imaging, for instance, the electric conductivity distribution in the human body carries information about the health condition of the patient, i.e., location of tumours, lung function and brain function, and hence an image of the conductivity is very useful for medical imaging \cite{holder2010a}. In other application domains such as Electrical Brain Stimulation it is important to have an accurate estimate of the brain's conductivity in order to compute the current density generated from the exterior \cite{saturnino2019a}. 

Electrical Impedance Tomography (EIT) is a fairly novel technology for medical imaging that aims at reconstructing a 3D image of a body's electrical conductivity from surface measurements of current and voltages through electrodes at the surface of the body. EIT is well-known that the inverse problem in EIT is notoriously ill-posed thus giving rise to low resolution images \cite{cheney1999a}. 
In recent years new ideas for conductivity imaging have been proposed that are often referred to as hybrid imaging. Broadly speaking, the idea of hybrid imaging is to utilize and control two separate, but coupled, physical phenomenon to obtain some extra information that makes the reconstruction problem more well-posed. 

One such method that has been subject to extensive research interest is acousto-electric tomography (AET) \cite{zhang2004a, WidlakScherzer2012}. In AET, one perturbs the electrical conductivity of the object by acoustic waves while conducting EIT measurements. The so-called acousto-electric effect allows one to recover first the internal electric power density and second the conductivity distribution. Mathematically, AET is well understood, but is has not yet matured as a technology. It is in fact questionable whether the measurable signals are strong enough for reconstruction purpose, or if the measurements are completely altered by unavoidable barriers like the omnipresent Johnson-Nyquist noise. The main goal of this paper is to clarify, using realistic models and parameters in a computational study, the feasibility of AET. In particular we need to give reasonable estimates of the Signal-to-Noise ratio (SNR) in an AET setup.

To formulate the problem mathematically we assume that the object of interest occupies some bounded domain $\Omega \subset \mathbb{R}^d,\, d = 2,3,$ with smooth boundary $\partial \Omega$. The object $ \Omega $ has the unknown isotropic electric conductivity $ \sigma \in L^\infty_+(\Omega)$ (bounded from above and below by positive constants). 
Through electrodes attached to the boundary, the normal current flux $f$ is controlled, and consequently an interior electric  potential $u$ is generated. When no interior sources or sinks of charge are present inside $\Omega,$ the potential is characterized by the generalized Laplace equation 
\begin{equation}
    \left\{\begin{aligned}
        -\nabla \cdot (\sigma(x)\nabla u(x)) &= 0, && x\in \Omega, \\
        \sigma(x)\partial_\nu u(x) &= f(x), && x \in \partial\Omega.
    \end{aligned}\right.
    \label{eq:u}
\end{equation}
Here $\nu$ denotes the outward unit normal vector to the boundary 
$\partial \Omega.$ We require that $f\in L^2_{\diamond}(\partial \Omega) = \{v \in L^2(\partial \Omega) : \int_{\partial \Omega} v\,\mathrm{d}s = 0\},$ i.e., the total input current vanishes. Then \eqref{eq:u} has a solution, and if we ground the potential by assuming $u|_{\partial \Omega} \in L^2_{\diamond}(\partial \Omega),$ $u\in H^1(\Omega)$  is uniquely determined. In EIT the aim is to reconstruct $\sigma$ from several measurements of $g= u|_{\partial \Omega}$ corresponding to a set of different input currents $f.$  

In AET the object is, in addition to the EIT measurement setup, penetrated by an acoustic wave that is generated by some source in the exterior of the body. Let $p$ denote the acoustic wave and $S$ the source function. We model $p$ by the scalar wave equation
\begin{equation}
    \left\{\begin{aligned}
        (\partial_t^2 - c^2(x) \Delta ) p(x,t) &= S(x,t), \quad &(x,t) &\in  \mathbb{R}^d \times \mathbb{R}_+  , \\
           p(x,0) = \partial_tp(x,0) &= 0,  &x &\in \mathbb{R}^d
    \end{aligned}\right.
    \label{eq:pressure}
\end{equation}
equipped with reasonable decay conditions. The sound speed $c$ and the source $S$ is considered fully known, and in that case $p$ is also known. When the acoustic wave travels through a material, the material is compressed and expanded. This results in a localized, time dependent change in the electrical conductivity, known as the acousto-electric effect. We denote the acoustically perturbed conductivity by $\sigma_p(x,t)$. 
A reasonable, yet simple model for $\sigma_p$ is the following \cite{alberti2018lectures}: 
\begin{equation}
    \sigma_{p}(x,t) = \sigma(x)(1 + \eta p(x,t)). \label{eq:sigmap}
\end{equation}
Here, $\eta > 0$ is the acousto-electric coupling parameter. We assume throughout that $\eta$ is known and constant. 

\begin{figure}[h!]
\centering%
    \subfloat[A][Phantom]{\includegraphics[height=0.30\textwidth]{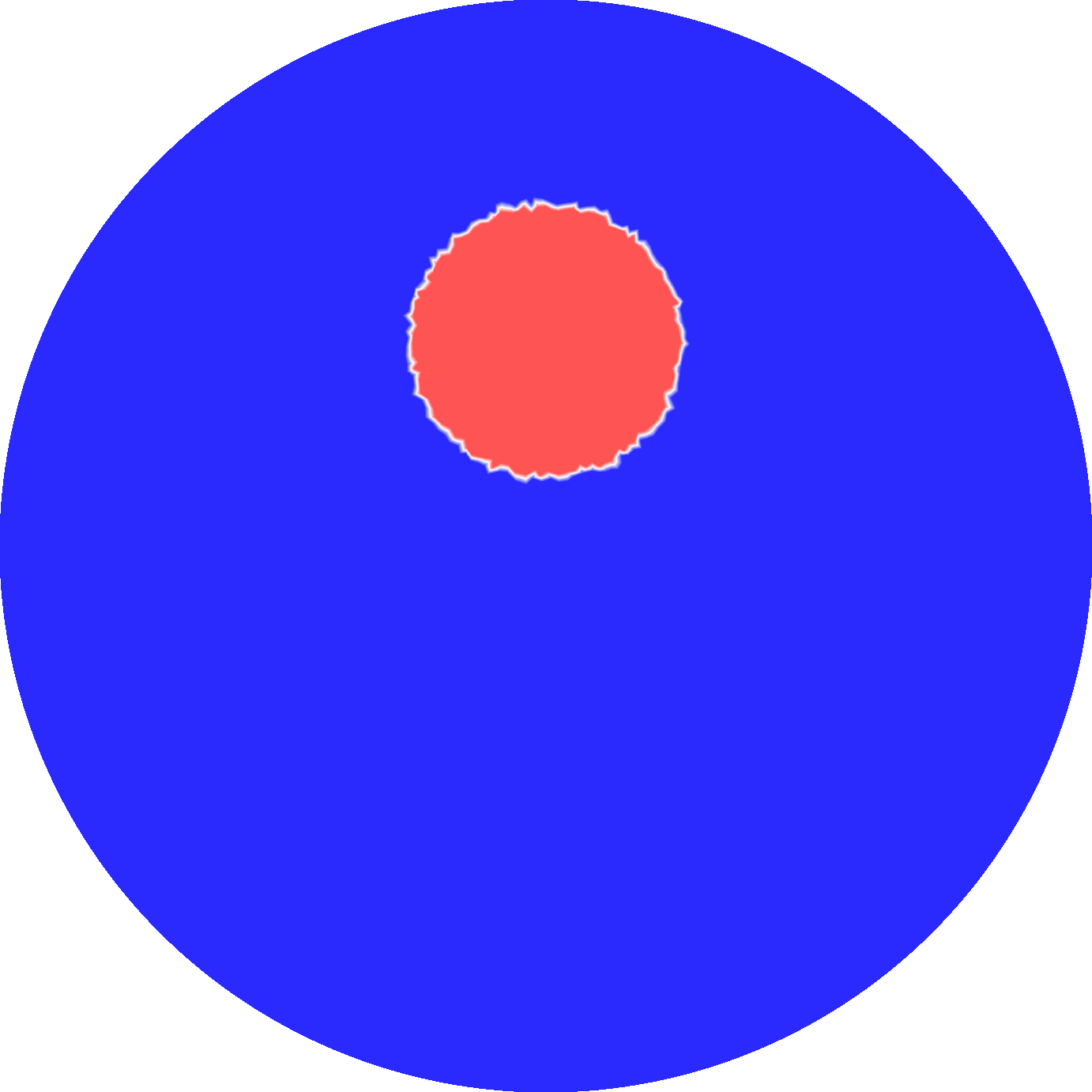}}%
    \hspace{0.75cm}%
    \subfloat[C][Acousto-electric effect]{\includegraphics[height=0.305\textwidth]{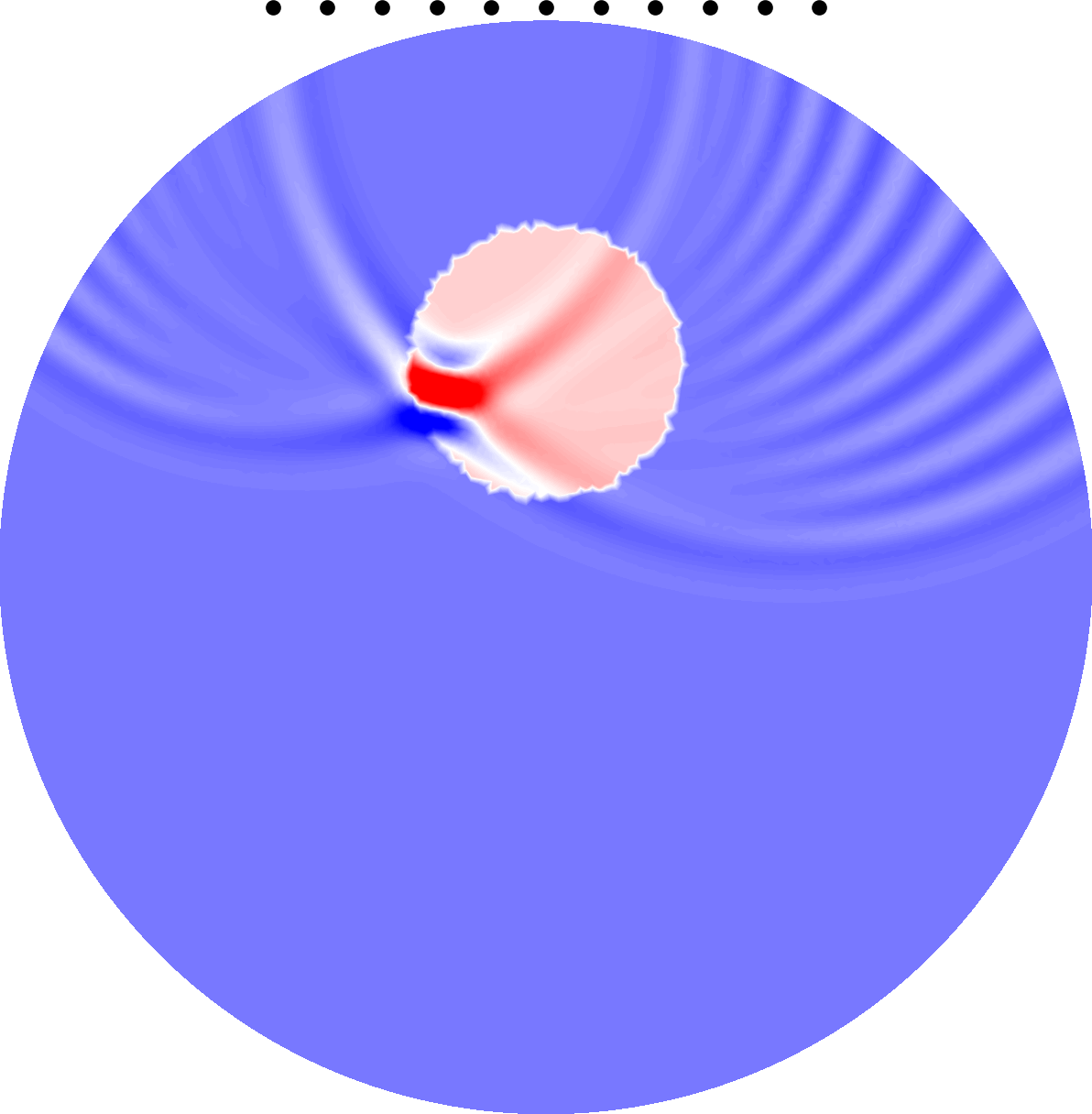}}%
    \caption{(a) Electric conductivity.
    (b) Perturbed conductivity due to the acousto-electric effect via a wave generated from transducers on the top; exaggerated for visibility.}
    \label{fig:phan_sigp}
\end{figure}

Figure \ref{fig:phan_sigp} illustrates the acousto-electric effect. The source $S(x,t)$ is situated outside the disk-shaped domain $\Omega$ and generates an acoustic wave. As the wave propagates through the domain, the conductivity is perturbed; an instantaneous image of the perturbed conductivity is seen to the right. 

The perturbation of $\sigma$ gives rise to a perturbation of the electric potential. We denote by  $u_p(x,t)$ this potential that for fixed $t \in \mathbb R_+$ solves the equation 
\begin{equation}
    \left\{\begin{aligned}
        -\nabla \cdot (\sigma_{p}(x,t)\nabla u_p(x,t)) &= 0, &&  x \in  \Omega,  \\
        \sigma_{p}(x,t)\partial_\nu u_p(x,t) &= f(x), && x \in \partial\Omega, 
    \end{aligned}\right.
    \label{eq:up}
\end{equation}
again with the convention $u_p|_{\partial \Omega} \in L^2_\diamond(\partial \Omega).$
The time-dependent perturbed  electric measurement is now $g_p = u_p|_{\partial \Omega} $, for $t \in [0,T]$ for some sufficiently large $T.$ 

In AET several boundary currents $f$ and wave sources $S$ are used, and the inverse problem is then to reconstruct the conductivity $\sigma$ from measurements of the corresponding voltages $g(x),g_p(x,t), \; x\in\partial \Omega, \; t \in [0,T].$ 

The typical approach to AET consists of two steps: First  the interior electric power density 
 \begin{align*}
    H(x) = \sigma(x) |\nabla u(x)|^2
\end{align*}
is reconstructed from the boundary measurements. Second, the conductivity $\sigma$ is obtained from $H$; this is the so-called quantitative step. In most literature on AET the first step is overlooked, and only the second step is considered. A novelty in this paper is that we, based on a carefully designed computational approach, consider both steps. We will see that the first step can be approached through a mildly ill-posed linear integral equation, while the second step can be approached as a (regularized) non-linear optimization problem. By numerical experiments we will demonstrate that AET is a feasible method for approximating the electrical conductivity in a body, for values of the electrical coupling constant in a range of values relevant for applications.

The signal in AET is generated by a second order effect driven by the  coupling parameter and the acoustic wave amplitude. We will, in a concentric situation, derive an explicit formula for the signal. The signal strength will then be compared to the ambient Johnson-Nyquist noise in the system; unfortunately this turns out to be relatively large. 

There is a vast literature on the mathematical and computational aspects of AET. The fundamental modelling originates from \cite{Ammari_Bonnetier_Capdeboscq_Tanter_Fink_2008}. In \cite{Kuchment_Kunyansky_2010,Kuchment_Kunyansky_2011} idealistic wave modelling was considered and an inversion scheme using the spherical Radon transform was derived. The reconstruction problem was considered in \cite{Capdeboscq_Fehrenbach_Gournay_Kavian_2009} using an optimal control approach. Several different linearization methods were compared in \cite{HoffmannKnudsen2014} giving rise to the analysis of artefacts in \cite{BalKnudsen:2018}. A numerical reconstruction method based on the Levenberg-Marquardt iteration was developed in \cite{Bal2012,li2021a} and methods using an explicit least squares optimization approach are found in \cite{adesokan2018fully,Roy-Borzi,jensen2018acousto,li2020a}; the limited angle problem was considered  in \cite{hubmer2018a,jensen2023a}. See also, e.g., \cite{monard2012a,MonardRim:2018,knudsen2023reconstructing}, for the case of anisotropic conductivities.

The outline of the paper is as follows: In Section \ref{sec:Inversion} we discuss the modelling of the inverse problem, explain the two step inversion procedure and it's numerical implementation. In Section \ref{sec:Parameters} the careful choice of physical parameters for the computational phantom is described, and the methods and tools for numerical simulation are introduced. Then in   Section \ref{sec:noise} we derive a formula for the signal strength in terms of the relevant parameters, compare this to an estimate of the Johnson-Nyquist noise, before described the chosen noise model for the computational experiments. Numerical experiments are carried out in Section \ref{sec:Numerical} before conclusions are drawn in Section \ref{sec:Conclusion}. 

\section{Inversion procedure and discretization}\label{sec:Inversion}

In this section we outline the procedure for the two-step inversion. Assuming knowledge of the acoustic fields, we first compute $H(x)$ using a truncated SVD decomposition for regularization. Next we solve for $ \sigma(x) $ using the method described in \cite{jensen2021sound}. We outline the approach for completeness.

\subsection{Step 1: From boundary measurements to power density}

\begin{figure}[!ht]
    \centering
    \includegraphics[width=1.0\textwidth]{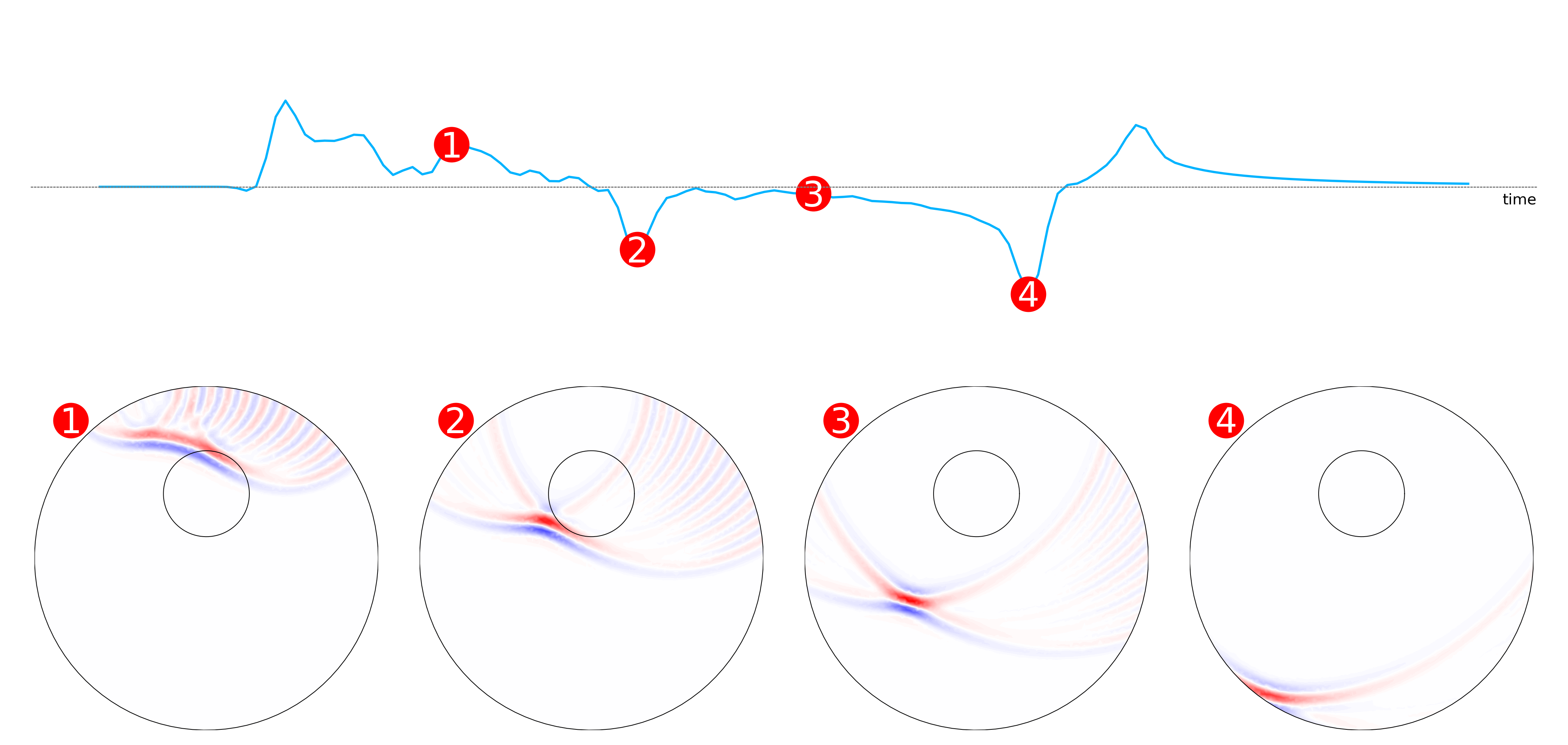}
    \caption{The top image contains the graph of the function $I(t)$ for the phantom seen in Figure \ref{fig:phan_sigp} corresponding to the illustrated propagating wave and the boundary condition $ f(x,y) = x$. The red vertical lines mark times corresponding to the four instant wave positions seen in the plots below.}
    \label{fig:I}
\end{figure}

Multiplying \eqref{eq:u} and \eqref{eq:up} with $u_p$ and $u$, respectively, subtracting, and integrating by parts yield the time series 
\begin{align}
    I(t) &= \langle f,g_p - g\rangle_{L^2(\partial\Omega)} \nonumber \\ 
    & = \int_{\partial\Omega}f(x)\left(g_p(x,t)-g(x)\right) \,\mathrm{d}S = -\eta\int_\Omega p(x,t)\sigma(x)\nabla u(x)\cdot \nabla u_p(x,t)\,\mathrm{d}x.
    \label{eq:I-form}
\end{align}
Physically, $I(t)$ measures at time $t$ the change in the required total electric power for maintaining the current flux $f$ at $\partial \Omega$ during the acoustic perturbation. Figure \ref{fig:I} illustrates how $I$ may behave as an acoustic wave travels through an object.  
Under the assumption that $\eta p $ is small, the approximation $u_p \approx u$ in $H^1(\Omega)$ is good (first order in $\eta$) and this yields the linear approximation
\begin{equation}
    I(t)  \approx -\eta \int_\Omega  p(x,t)\sigma|\nabla u(x)|^2\,\mathrm{d}x = -\eta \int_\Omega  p(x,t)H(x)\,\mathrm{d}x.
    \label{eq:inteq2} 
\end{equation}
We will work in this linear regime. By controlling the source $ S $ (see \ref{sec:Numerical} below) we generate a sequence of waves each with a different focus point in $\Omega$. We represents all of these waves together by $ p(x,t) $ concatenated sequentially in time. To each choice of boundary condition $ f_i $, $i = 1,\dots, N_f $, there is a power density $ H_i $ related to corresponding time signal $I_i$ by
\begin{align}
    I_i(t) = -\eta\int_\Omega p(x,t)H_i(x)\,\mathrm d x, \quad 1 \leq i \leq N_f.
\end{align}
Using a finite element basis $ \varphi_n(x) $, $ n = 1,\dots, N_\varphi $, we discretize in space and time, representing 
\begin{align}
    H_i(x) &= \sum_{n=1}^{N_\varphi} H_{i,n}\varphi_n(x), \quad \mathbf{H}_i = \left[ H_{i,1}, \dots, H_{i,N_\varphi} \right]^T, \\
    p(x,t_j) &= \sum_{n=1}^{N_\varphi} p_{j,n}\varphi_n(x), \\
    \mathbf{I}_i &= \left[I_i(t_1), \dots, I_i(t_{N_t})\right]^T. \label{eq:I-vec}
\end{align}
Let $ \mathbf{M} = (a_{m,n})_{m,n} $ denote the mass matrix with elements $ a_{m,n} = \int_\Omega\varphi_m\varphi_n\,\mathrm dx $ and $ \mathbf{P} = (p_{j,m})_{m,j} $ the wave-matrix in which column $ j $ contains the finite elements coefficients of the wave at time $ t_j $. Then we discretize the integral equation \eqref{eq:inteq2} by the matrix $ \mathbf{K} = -\mathbf{P}^T\mathbf{M} $ acting on $ H $
\begin{align}
    \eta \mathbf{K}\mathbf{H}_i = \mathbf{I}_i, \quad \text. \label{eq:discrete-problem-H}
\end{align}

We then solve \eqref{eq:discrete-problem-H} by computing an SVD decomposition and regularize the solution by truncation.

\subsection{Step 2: From power density to conductivity}
The problem of reconstructing $ \sigma $ from the interior power density data $ \mathbf{H} $ has been explored in a variety of scenarios, see for example \cite{Ammari_Bonnetier_Capdeboscq_Tanter_Fink_2008,
Gebauer_Scherzer_2008,Kuchment_Kunyansky_2011,HoffmannKnudsen2014,BalKnudsen:2018}. We approach this problem by minimizing the functional
\begin{align} \label{eq:J}
    \min_{\sigma} \mathcal J(\sigma) = \sum_{i=1}^{N_f}\|H_i(\sigma) - z_i\|_{L^1(\Omega)} + \beta|\sigma|_{\mathrm{TV}},
\end{align}
where $ z_i(x) = \sum_{n=1}^{N_\varphi}\mathbf{H}_{i,n}\varphi_n(x) $ is the reconstructed data from step 1. The $ L^1 $ data-fidelity terms depends on the assumptions on the regularity of $ \sigma $. In particular, if $ \sigma $ has discontinuities, standard regularity theory only guarantees $ L^1 $-regularity of $ H $; with $ L^{1+\varepsilon} $-regularity for sufficiently small discontinuities, see \cite{jensen2018acousto}.

For reconstruction we use the algorithm described in \cite{jensen2018acousto}. That is, we consider the following weighted quadratic problem arising from a linearization of \eqref{eq:J},
\begin{align} \label{eq:Jquad}
    J_{\sigma,\kappa'}(\kappa) = \frac12\sum_{i=1}^{N_f}\int_\Omega w_i(\sigma,\kappa')|H_i'(\sigma)[\kappa] - z_{i,\sigma}|^2\,\mathrm dx + \frac\beta2\int_\Omega w_0(\sigma,\kappa')|\nabla (\sigma+\kappa)|^2\,\mathrm dx,
\end{align}
where $ z_{i,\sigma} = z_i - H_i(\sigma) $ and the weights are
$$ w_i(\sigma,\kappa') = |H_i'(\sigma)[\kappa'] - z_{i,\sigma}|_{\epsilon}^{-1}, \quad \text{for $ 1 \leq i \leq N_f$,} \quad w_0(\sigma,\kappa') = |\nabla(\sigma+\kappa')|_\epsilon^{-1}, $$
with $ |\cdot|_\epsilon \equiv \sqrt{|\cdot|^2 + \epsilon^2} $.

To solve \eqref{eq:J} we initialize $ \kappa' $ as zero, compute the weights $ w_i $, and successively do partial minimization of \eqref{eq:Jquad} to update $ \kappa' $ and thus the weights. After a few rounds of this we update $ \sigma := \sigma + \kappa' $ and restart the process. We do this following the steps outlined in \cite{jensen2021sound} with the small modification that we in each step also project $ \sigma := \sigma + \kappa $ to the interval $[10^{-2},\infty)$ since $ \sigma $ cannot physically become negative or too close to 0, which would imply complete electrical isolation.

\section{Parameters and numerical modeling}\label{sec:Parameters}
In this section we describe the parameters and phantom used in the numerical experiments. Furthermore, we give a detailed description of the modeling and numerical simulation of the acoustic sources and fields, the electrical potential, and the measurements.

\subsection{The computational phantom}
We stick to a simple two dimensional model for clarity insight and computationally efficiency. The approach can in a straightforward way be generalized to more complex domains in two and three dimensions. 

The computational phantom draws inspiration from breast cancer imaging: We assume a circular domain modelling a cross section of the breast. The domain consists of homogeneous material, normal breast tissue. In the domain is a small region that contains different material, the cancerous tissue. The phantom is illustrated in Figure \ref{fig:phantoms} (a). 

For reference the domain $\Omega = \{ x \in \mathbb{R}^2 : |x| < R \}$ with $R=0.04\,\si{m}.$ The embedded region is also circular centered at $(0, 0.015)$m with radius $0.01$m.
 
\subsubsection*{Electrical conductivity and boundary current density}
Many authors have reported results on the electrical conductivity of human tissue, and from \cite{joines1994,hesabgar2017dielectric,bidhendi2014ultra} one infers that the electrical conductivity differs by a factor in the range 5 to 15 between healthy and cancerous breast tissue and experimental values for $\sigma$ are reported to lie in the range $0.01$ to $1\,\si{S}/\si{m}$, see, e.g.,  \cite{joines1994,cherepenin20013d,barber1984applied,hesabgar2017dielectric}. In our experiments, we choose a high contrast conductivity phantom taking values in the range $0.1$ to $1.0\,\si{S}/\si{m}.$

Further, the maximum allowed input current is $f_\text{max}=1\,\si{mA}$ \cite{cherepenin20013d}. We enforce this by requiring that 
$$ \int_{\partial \Omega} f^+ \mathrm{d}s \leq f_\text{max}, \quad \text{where} \quad f^+(x) = \operatorname{max}\{f(x),0\}.$$

\subsubsection*{Acousto-electric coupling and acoustic pressure}
The value of the acousto-electric coupling constant $\eta$ is of great importance. The coupling is known to be weak, and therefore the conductivity perturbation is also small. Via the indirect effect the signal $I$ is therefore very small. In \cite{li2012measuring}, $\eta$ in a rabbit heart is found to be approximately $0.041\,\si{MPa^{-1}}$, i.e., $\eta \approx 4.1 \times 10^{-8}\,\si{Pa^{-1}}$, while in \cite{song2016image}, values $\eta$ in $0.9 \%\,\si{NaCl}$ solution is reported to be of the magnitude $10^{-9}\,\si{Pa^{-1}}$. In the literature we could not find measured values of $\eta$ for breast tissue, but the mentioned values indicates a range of interest. We chose to perform our numerical experiments for $\eta = 10^{-8}.$ 

We choose the maximal amplitude of the acoustic pressure $p$ to be $p_\text{max} = 1.5\,\si{MPa}$, in accordance with clinical standards \cite{demi2014basics}, and assume a constant wave speed of $c_0 = 1500\,\si{m}/\si{s}$ \cite{demi2014basics}. 

\begin{figure}[!ht]
\centering
    \subfloat[A][Phantom]{\raisebox{1.5mm}[0pt][0pt]{\includegraphics[height=5.15cm]{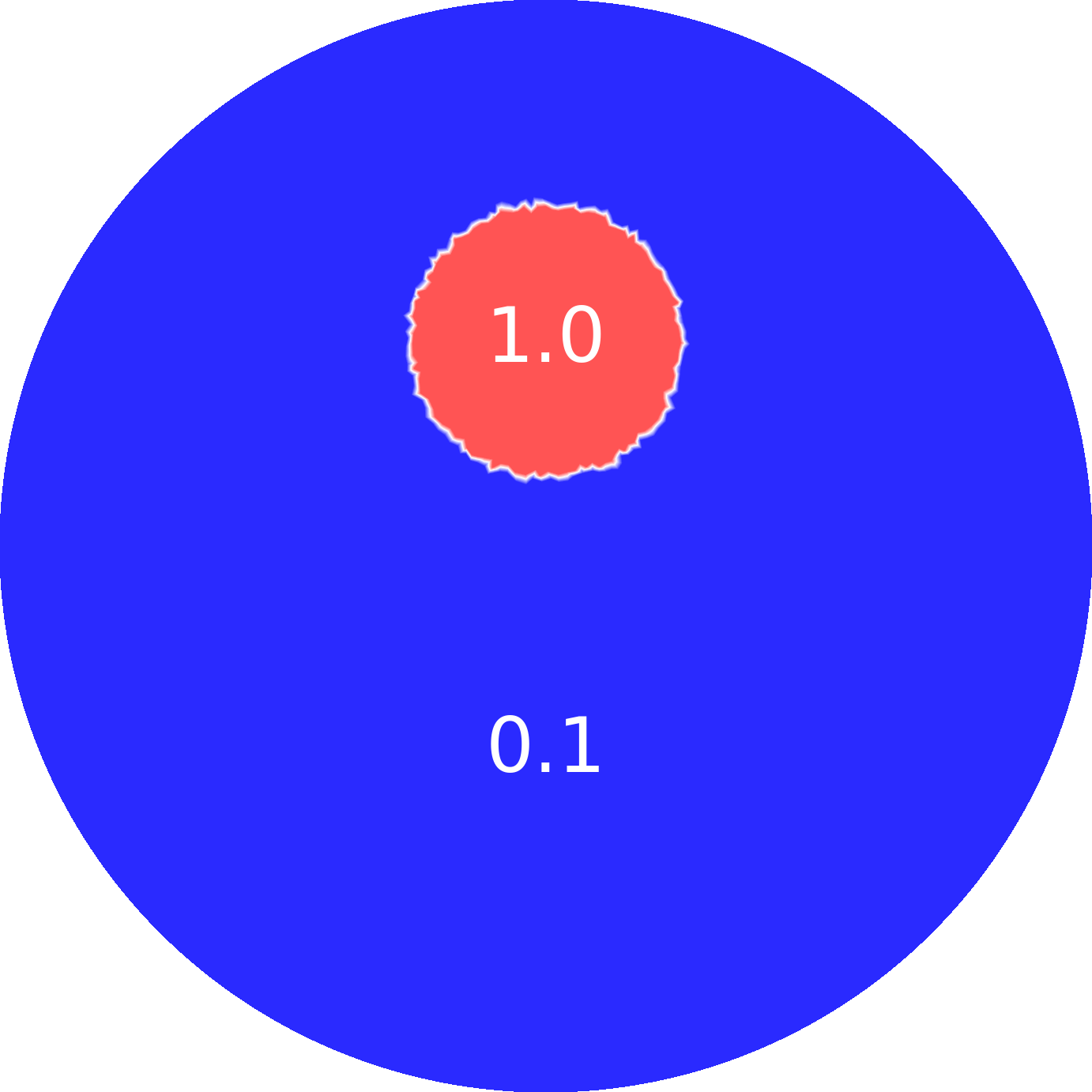}}}%
    \hspace{1cm}%
    \subfloat[B][Focus points and transducers]{\includegraphics[height=5.5cm]{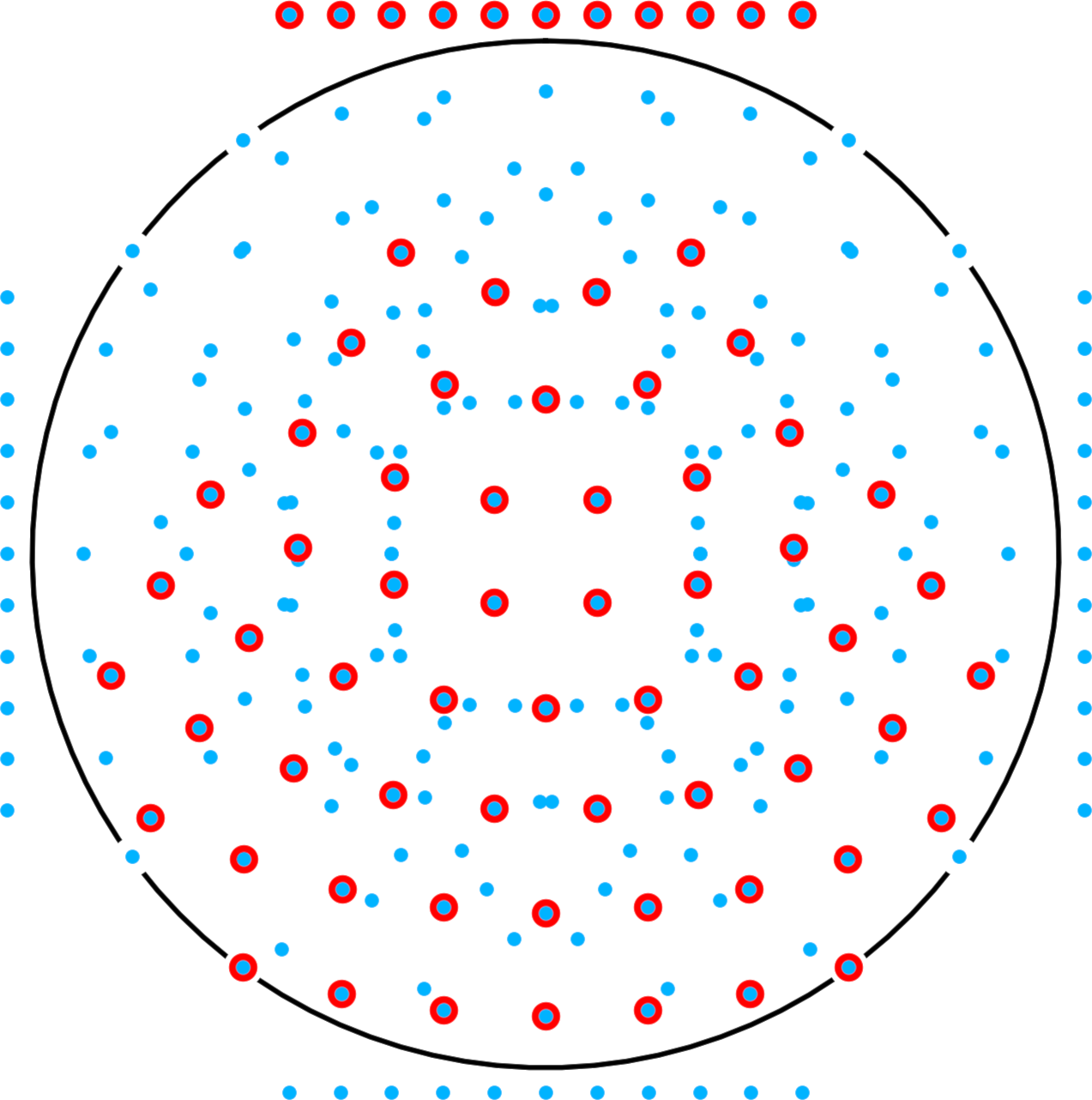}}

    \caption{(a) The phantom conductivity. The conductivity is 1.0 in the disc inclusion and the background is 0.1. (b) The points where the wave fronts gets focused and the transducers. Each transducer focus waves in a subset of the focus points. The red-ringed transducer at the top focuses in the cone of red-ringed points in the domain. The point distribution corresponding to each transducer is rotationally symmetric. In simulations there are no gaps between transducer arrays and the domain.}
    \label{fig:phantoms}
\end{figure}

The acoustic wave is generated by a transducer array situated outside the phantom. The transducer array consists of equidistant individual point sources along a line segment. Each of the point sources excites one period of sine-wave with a frequency of $250 \,\si{kHz}.$ By controlling the time delay of the individual transducers, we can simulate many different wave patterns. We choose to work with focusing waves, i.e., waves that at a given time approximately concentrates at a certain point in the domain, see Figure \ref{fig:phan_sigp} and \ref{fig:I}. We use 4 equidistant transducer positions around the boundary (above, below, left, right) and at each position we have a unique set of 58 focus points as illustrated in Figure \ref{fig:phantoms}(b): blue points indicate all focus points, red points the focus for the above positioning of the array. In Table \ref{tbl:parameters}, we summarize our parameter choices. 

\begin{table}[t]
    \centering
    \setlength\extrarowheight{2.5pt}
    \caption{Model parameters} \label{tbl:parameters}
    \begin{tabular}{| l | l | l | l |  }
    \hline
    Parameter & Symbol & Value(s) & Unit \\ \hline
    \hline 
    Conductivity & $\sigma$ & $0.1-1.0$ & \si{S/m} \\ \hline
    Total boundary current flux & $ f_{\text{max}} $ & $1\times 10^{-3}$ & \si{A} \\ \hline
    Acousto-electric coupling constant & $\eta$ & $10^{-8} $ & \si{Pa^{-1}} \\ \hline
    Max. acoustic pressure & $p_\text{max}$ & $1.5 \times 10^6$ & \si{Pa} \\ \hline
    Acoustic wave speed & $c_0$ & $1500$ & \si{m/s} \\ \hline
    Domain radius & $R$ & $0.04$ & \si{m} \\ \hline
    \end{tabular}
\end{table}

\subsection{Simulation of the forward problem}
\subsubsection*{Acoustic field and source terms}

The expression \eqref{eq:inteq2}  suggests that one could either choose a set of acoustic fields that approximate some basis (for e.g. $L^2(\Omega)$), or a set of fields that approximate a Dirac distribution at a sufficiently large set of points in $\Omega$. In this work we have used different focused acoustic waves to sample the the power density throughout the domain.

We simulate the acoustic field on a square finite difference grid containing $\Omega$. \kwave{} is a \matlab{} package that uses a $k-$space pseudo-spectral method to effectively and accurately solve the wave equation. To avoid the problem of having to prescribe boundary conditions on the finite computational domain, \kwave{} uses a so-called perfectly matched layer that absorbs the wave at the domain boundary.
For the source terms $S_j$ of our acoustic field, we use a simplified transducer model. We model the action of the transducer as a linear array of $N_p = 11$ time dependent volume point sources. We assign to the \kwave{} source function a discrete set of $N_p$ grid points on a line segment of length $ \ell = 4 \si{cm} $ at the boundary of the domain; e.g. see Figure \ref{fig:phantoms}(b).

Further, we use the focus functionality to delay the individual point source signals and orient the wave in various directions. See the \kwave{} function \texttt{focus.m} and the \kwave{} manual \cite{treeby2010k} for further details and explanations on this, and on how \kwave{} solves the wave equation.

We simulate acoustic fields with the above parameters on a uniform 350-by-350 grid with physical dimensions $12\,\si{cm}\times 12\,\si{cm}$. We place the transducer tangentially to the boundary at 4 equidistant points along the boundary. At each position we focus the acoustic wave at a set of 58 points distributed in a cone originating at the transducer center. The 4 positions are illustrated in Figure \ref{fig:phantoms}(b) where the cone of points corresponding to the top-most transducer location are also highlighted; we describe how this cone of points was chosen in the next section. We sample the waves equitemporally at $ N_t = 141 $ timestamps in the range $ 0 $ to $ T $, where $ T $ is a uniform upper bound to let the wavefront propagate through the domain. $ T $ is set by \kwave{}. The acoustic fields $ x \mapsto p(x,t_j) $ are then interpolated to the triangulated mesh for the domain $ \Omega $.

\subsubsection*{Focus point geometry}
We construct the focus points by the following procedure: From the transducer location $ q $ we place a cone $ C $ with axis towards the center of $ \Omega $ and angle $ \theta $ between axis and cone side. We choose a minimum distance $ r_{\text{min}} $ between focus points and the transducer and a focus point distance $ \Delta r $. 

Along each arc of the circle with center $ q $ and radius $ r(k) = r_{\text{min}} + k\Delta r $, $ k = 1,\dots $, which has non-empty intersection with $ C \cap \Omega $, we distribute the maximum number of points, with arc-distance $ \Delta r $ between them, symmetrically across the cone-axis.

For our simulations, illustrated in Figure \ref{fig:phantoms}(b), we used the values $$ r_{\text{min}} = \frac{R}{2}, \quad \Delta r = \frac{R}{5}, \quad \theta = \tan^{-1}\left(\frac{3}{4}\right). $$

\subsubsection*{Electric potential measurements}
The electric potentials are computed by solving the PDE \eqref{eq:u} using the open source computing framework \fenics{} \cite{fenics}. 

The discretization is done using a $\mathcal P_1 $ FEM-basis over a 16,617-node triangulated mesh for $\Omega$. For each timestamp $t_j$ and corresponding acoustic field $ p(x,t_j) $ we compute the perturbed conductivity $\sigma_p(x,t_j)$ according to \eqref{eq:sigmap}. The PDE is then solved for the potential field $ u_p(x,t_j) $ and we take the inner product of the difference between $ u_p(x,t_j) $ and the static field $ u(x) $ with the Neumann-boundary $ f(x) $ on the boundary $ \partial\Omega $ to obtain the derived data $ I(t_j) $ as outlined by \eqref{eq:I-form}. This is done for each of three different boundary conditions:
\begin{align}
    f_1(x,y) = x, \quad f_2(x,y) = y, \quad\text{and}\quad f_3(x,y) = \frac{x+y}{\sqrt{2}}, \label{eq:boundary-conditions}
\end{align}
leading to the data $ I_i(t_j) $ corresponding to $ f_i $, each set bundled in vectors $ \mathbf I_i $ as outlined in \eqref{eq:I-vec}.

\section{Signal and noise magnitude}\label{sec:noise}

As described above, the acousto-electric effect is weak, and the changes in the voltage potential induced by the acoustic perturbation are very small. This makes it challenging to separate the signal from the noise in the boundary measurements, and can be a serious problem for AET. 

For weak electrical signals like that of AET, there is a source of noise that cannot be eliminated. This is the so-called thermal noise. As it is very complicated to accurately model such noise in a complex system like AET (the modelling must also involve the measurement equipment), we choose a different strategy. We instead derive a simple estimate for the AET signal magnitude $I(t)$ as a function of the central parameters. We then test our inversion method for various levels of noise (relative to $I(t)$). This then allow us to give a quantitative bound on the magnitude of the Johnson-Nyquist noise for which AET is feasible, given a specific set of parameters.  

\subsection{AET signal magnitude in a radial geometry}

We first derive a simple estimate of the difference between a acoustically perturbed and unperturbed AET-measurement. As before, let $\Omega$ be a disc in $\mathbb{R}^2$ of radius $R$ and centered at the origin, and $\Omega_p = \{x \in \Omega : |x| < R_p \leq R\} $
be the part of $\Omega$ perturbed by the acoustic wave $p(x,t)$.  Let $\sigma_0 > 0$ be a constant conductivity and let $\sigma_p = \sigma_0(1 + \varepsilon\chi_{\Omega_p}(x)) $, where $\chi_{\Omega_p}(x)$ is the characteristic function for $\Omega_p$ and $\varepsilon = \eta p_{\text{max}}$ is small and dimensionless. For a boundary current $ f \in L^2(\partial\Omega) $ let $ u_0 $ be the solution of \eqref{eq:u} for $ \sigma := \sigma_0 $, and $ u_p $ the solution of \eqref{eq:up} with $ \sigma_p $ as defined in the text immediately above. As in \eqref{eq:I-form} we form $ I(t) = \langle f, g_p - g_0\rangle_{L^2(\partial\Omega)} $ where $ g_p = u_p\vert_{\partial\Omega} $ and $ g_0 = u_0\vert_{\partial\Omega} $.

We are interested in quantifying $I$ and $g_p - g_0$ in terms of the involved parameters, especially the dependence on the coupling constant $\eta$. Modifying the approach in \cite{mueller2012linear} (p. 166-168) to a Neumann-to-Dirichlet situation, we find that 
$$ g(\theta) = \sum_{n\in\mathbf{Z}} \frac{\hat{f}_n}{|n|\sigma_0}R \mathrm{e}^{in\theta},  \quad g_p(\theta) = \sum_{n\in\mathbf{Z}} \frac{\hat{f}_n}{|n|\sigma_0}R\mathrm{e}^{in\theta} \frac{2 + \varepsilon(1-\left(R_p/R\right)^{2|n|} )}{2 + \varepsilon(1+\left(R_p/R\right)^{2|n|} )}, \quad 0 \leq \theta < 2 \pi,$$
where $\hat{f}_n$ is given by $ \hat{f_n} = (2\pi)^{-1}\int_0^{2\pi} f(\theta) \mathrm{e}^{-in \theta} \mathrm{d}\theta.$
We now get that 
\begin{align*}
  g_p(\theta) - g(\theta) = \sum_{n\in \mathbf{Z}}\frac{\hat{f}_n}{|n|\sigma_0}R\mathrm{e}^{in\theta}\left( -\frac{2\varepsilon \left(R_p/R\right)^{2|n|}}{2 + \varepsilon(1+\left(R_p/R\right)^{2|n|})} \right). 
\end{align*} 

The function $f$ is the current density of the input current. Setting $f = C_f \mathrm{e}^{i\theta}$ we obtain the requirement that 
\begin{equation} 
    \int_{\partial \Omega} \max(\text{Re}(f),0 )\mathrm{d}\theta \leq f_{\text{max}}, \label{eq:f-max}
\end{equation}
 which leads to $C_f = \frac{f_{\text{max}}}{2 R}$. Also, this choice for $ f $ yields 
\begin{equation}
    g_p(\theta) - g(\theta) = \frac{f_{\text{max}} }{4 \pi \sigma_0} \mathrm{e}^{i\theta} \left(-\frac{2\varepsilon \left(R_p/R\right)^{2}}{2 + \varepsilon(1+\left(R_p/R\right)^{2})} \right). 
\end{equation}
We define the signal amplitude $A$ as the coefficient to $ e^{i\theta} $: 
$$ 
    A = \frac{f_{\text{max}}}{4 \pi \sigma_0}  \left(-\frac{2\varepsilon \left(R_p/R\right)^{2}}{2 + \varepsilon(1+\left(R_p/R\right)^{2})} \right).
$$
For $\varepsilon \ll 1$ and using the boundary inner product formula for $ I $ \eqref{eq:I-form} and the above choice for $f$ and \eqref{eq:f-max}, we obtain 
$$ |A| \approx \frac{f_{\text{max}}}{4 \pi \sigma_0} \left(\varepsilon \left(R_p/R\right)^{2} \right) \quad \text{and } \quad |I| \approx \frac{f^2_{\text{max}}R}{2 \sigma_0} \left(\varepsilon \left(R_p/R\right)^{2} \right). $$

We use the parameters in Table \ref{tbl:parameters} and pick the perturbation radius as one tenth of the domain radius $ R_p = \frac{R}{10} $. Recall that $ \varepsilon = \eta p_{\text{max}} $. With these values we find that
$$
    |A| \approx 1.2 \times 10^{-7} \quad \text{and} \quad |I| \approx 3.0 \times 10^{-11}.
$$

The dependency of $ I $ on the parameters is interesting. Both $ p_{\text{max}} $ and $ \eta $ enters in $ \varepsilon $ and thus the magnitude $ |I| $ linearly, whereas $ f_{\text{max}} $ enters quadratically. We can hence increase $|I|$ through $ p_\text{max}$ and $ f_\text{max}, $ however these quantities are constrained by health considerations in particular in medical imaging applications.

\subsection{Measurement noise and errors}

The signal amplitude in AET is very small as seen above. At such small scale   we need to consider the presence of ambient thermal noise, i.e., noise due to the motion of the free electrons in the material itself. Such noise is called Johnson-Nyquist noise and cannot be decreased by improving the measurement technology and system. Usually, Johnson-Nyquist noise is modelled as a stochastic process independent of the other electrical activity going on. In a simple resistor system, it is an additive Gaussian white noise process, and the root-mean-square of the voltage signal of a thermal noise process is given by the formula 
\begin{equation}
    V_{\textup{RMS}} = \sqrt{4k_BTR\,\Delta\!f},
    \label{eq:johnson-nyquist}
\end{equation}
where  $k_B = 1.380649\times 10^{-23}$  is the Boltzmann constant, $ T $ the absolute temperature of the resistor, $ R $ the resistance and $ \Delta\!f $ the effective bandwidth of the measurement signal. We refer to \cite{gillespie1996mathematics, gillespie1998a} for more details regarding Johnson-Nyquist noise. 

We need to assume some reasonable sizes for the involved quantities; we are mostly interested in the orders of magnitude rather than exact numbers. The temperature $ T = 293.15\, \si{K}\approx 20 {}^\circ C,$ that is living room temperature. The effective bandwidth is fixed to $ \Delta f = 300 $ kHz; the quantity can be estimated via the Fourier transform of signal. Finally, the
resistance $R$ is in an EIT system in the range $ 1\,\si{k\Omega} $ to $ 100\,\si{k\Omega}.$ 
Putting these numbers into \eqref{eq:johnson-nyquist} gives the estimate
$$ 
     V_{\textup{RMS}} = \sqrt{4k_BTR\Delta f} \in (2.20\,\mu\si{V},\, 22.0\,\mu\si{V}).
$$
Let's for the moment fix $V_{\textup{JN}} = 10\;\mu$V. The impact in the $I$ signal can then be estimated by
\begin{align*}
    I_{\textup{JN}} &= \int_{\partial\Omega} V_{\textup{JN}} \;C_f\,\mathrm{d}\theta\\
    &=\int_{\partial\Omega} \overline{V}_{\textup{RMS}} \frac{f_\textup{max}}{2R}\,\mathrm d\theta = \pi \cdot 10\times 10^{-6} \cdot {10^{-3}} = \pi 10^{-8}\,\si{W}.
\end{align*} 
Comparing the estimated Johnson-Nyquist noise $ I_{JN} = \pi \times  10^{-8} \si{W} $ to the calculated signal $|I| \approx 3 \times 10^{-11} \si{W}$ is somewhat discouraging. By repeating the measurement one could improve the noise by a decade or two and even possibly bring the signal and noise at par. We will in the next section numerically model the noise and work with an SNR of 0 dB and below; equivalently a relative noise level of 100$\%$ and more.

Let's remark that the electrical system involved in AET is immensely more complicated than the simple circuit for which formula \eqref{eq:johnson-nyquist} is derived. It involves biological matter, electrical measurement apparatus and acoustic perturbations, and it is unknown where in the system the critical contribution to the noise will appear (e.g., in the domain, in the electrodes, in the wires from the electrodes, etc.). It is beyond the scope of this investigation to try to precisely quantify the magnitude of the noise in a more realistic AET system.  

\subsection{Noise modelling} \label{noise}
Modelling the noise in AET there are a myriad of potential sources of stochasticity including, but not limited to, the exactness of the applied boundary current, the precision of boundary potential measurements, the knowledge of the wave (could be related to precision of transducer positions, transducer timing, modelling limitations, wave speed, etc.) and exactness of timestamps for measurements etc.

We choose a simple and practical approach to noise modelling in line with what is usually done in inverse problems modelling. We assume that the  noise enter the model as additive Gaussian noise on  the signal $I$.

Concretely, let $\mathbf{I}_i$ be the sampled true measurement of $I(t),$ see \eqref{eq:I-vec}. We then add Gaussian noise by first generating the Gaussian vector $\mathbf e \sim \mathcal N(0, \mathbf{Id})$ and forming
$$
    \widetilde{\mathbf I}_i = \mathbf I_i + s\|\mathbf I_i\|_2\left(\frac{\mathbf e}{\|\mathbf e\|_2}\right).
$$
Here $ s $ indicates the relative noise level. In accordance with the analysis above, we will consider $s \in \{0,1,5,10\},$ i.e., relative noise of $0\%, 100\%, 500\%$ and $1000\%.$




\section{Validation of the numerical method and feasibility of AET}\label{sec:Numerical}
In this section we put the above theory and numerical methods to work on the measurements generated by the phantoms in Section \ref{sec:Parameters}. We first demonstrate that we can reconstruct the power density, and then apply our method to simulated measurements for a range of parameters and noise levels. 

\subsection{Reconstructing the power density}
We pick the finite element basis $ \varphi_n $, $ n=1,\dots,N_\varphi $, from Section \ref{sec:Inversion} as a $ \mathcal{P}_1 $ basis on an unstructured 6,065-node triangulated meshing for $ \Omega $. With this we construct the discretized integral operator from equation \eqref{eq:discrete-problem-H} with the the discretized waves $ p(x,t) $ temporally sampled in the timestamps corresponding to the timestamps of the given noisy data $ \widetilde{\mathbf I}_i $. 

We solve the equation \eqref{eq:discrete-problem-H} for $ \mathbf H_i $ using a truncated singular value decomposition, though not straight forward from $ \mathbf K $. Instead we rewrite the equation on the form
\begin{align}
    \widetilde{\mathbf{K}}\widetilde{\mathbf{H}}_i = \widetilde{\mathbf{I}}_i
\end{align}
where $ \widetilde{\mathbf K} = \eta \mathbf K\mathbf L^{-T}$, $\widetilde{\mathbf{H}}_i = \mathbf L^T\mathbf H_i $ and $ \mathbf L $ is the Cholesky factor of the mass matrix $ \mathbf M = \mathbf L\mathbf L^T $.

We the take a singular value decomposition $ (\mathbf U, \boldsymbol{\Sigma}, \mathbf V^T) $ of $ \widetilde{\mathbf{K}} $ and do a truncated reconstruction of $ \widetilde{\mathbf{H}}_i $. To this end we construct the truncated pseudo-inverse $ \boldsymbol{\Sigma}_k^\dagger $ and compute
\begin{align*}
    \widetilde{\mathbf{H}}_i = \mathbf V \boldsymbol{\Sigma}_k^\dagger \mathbf U^T\widetilde{\mathbf I}_i,
\end{align*}
where $ k $ is the truncation threshold. We can then recover $ \mathbf H_i $ from that result by inverting $ \mathbf L^T $.

This is done for each $ \mathbf H_i $ corresponding to each of the boundary condition in \eqref{eq:boundary-conditions}.

\begin{remark}
The reason for this restructuring of \eqref{eq:discrete-problem-H} before the singular value decomposition is that the resulting decomposition corresponds to a decomposition for the wave matrix $ \mathbf P^T $ such that if we take the matrix $ \widetilde{\mathbf{V}} = \mathbf L^{-T}\mathbf V $ then it is the right singular vectors for $ \mathbf P^T $ in a space where the inner product is $ \langle \cdot, \cdot\rangle_{\mathbf{M}} $, that is in the discretized $ L^2(\Omega) $ inner product; so $ \widetilde{\mathbf{V}}^T\mathbf M\widetilde{\mathbf{V}} = \mathbf{Id} $. Hence the singular vectors of $ \widetilde{\mathbf V} $ have a meaningful visualization on our finite element mesh.
\end{remark}

\begin{figure}[!ht]
\centering
    \newcommand{\scale}{0.29}
    \subfloat[A][$k = 1500$]{\includegraphics[height=\scale\textwidth]{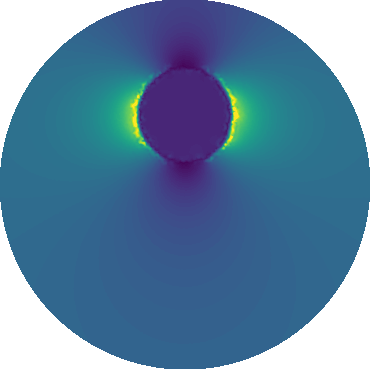}}~\subfloat[A][$k = 1500$]{\includegraphics[height=\scale\textwidth]{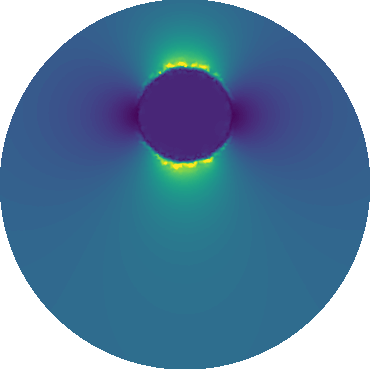}}~\subfloat[A][$k = 1500$]{\includegraphics[height=\scale\textwidth]{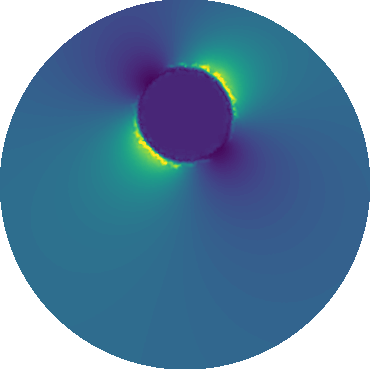}}~
    \includegraphics[height=\scale\textwidth]{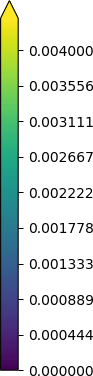}%
	\caption{True power densities.}
    \label{fig:H-true}
\end{figure}

\begin{figure}[!ht]
\centering
    \newcommand{\scale}{0.29}
    \subfloat[A][$k = 1500$]{\includegraphics[height=\scale\textwidth]{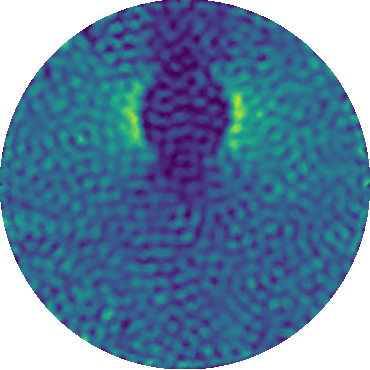}}~\subfloat[A][$k = 1500$]{\includegraphics[height=\scale\textwidth]{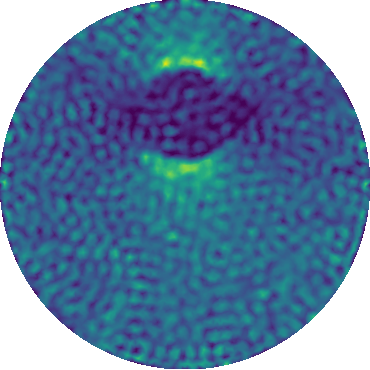}}~\subfloat[A][$k = 1500$]{\includegraphics[height=\scale\textwidth]{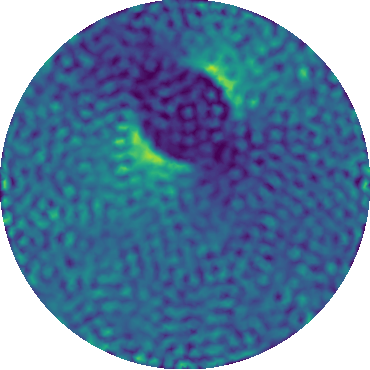}}~
    \includegraphics[height=\scale\textwidth]{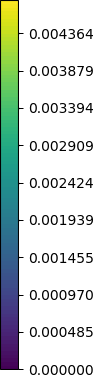}%
    \\%
    \subfloat[B][$k = 1200$]{\includegraphics[height=\scale\textwidth]{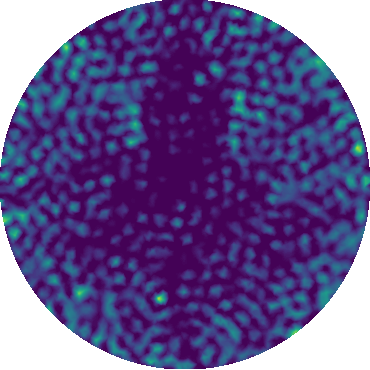}}~
    \subfloat[B][$k = 1200$]{\includegraphics[height=\scale\textwidth]{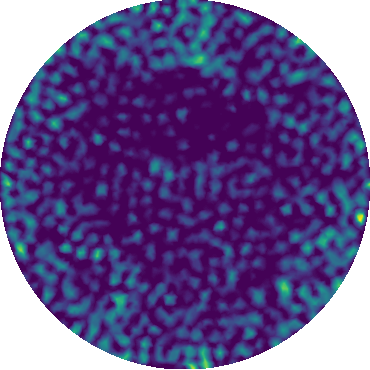}}~
    \subfloat[B][$k = 1200$]{\includegraphics[height=\scale\textwidth]{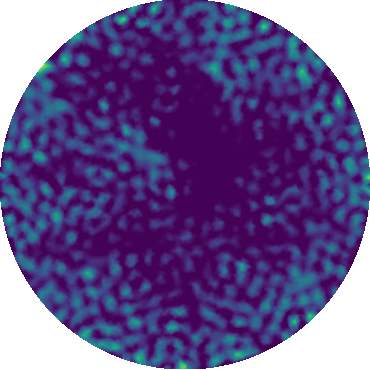}}~
    \includegraphics[height=\scale\textwidth]{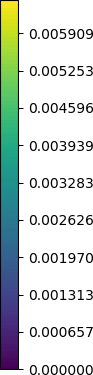}%
    \\%
    \subfloat[C][$k = 1100$]{\includegraphics[height=\scale\textwidth]{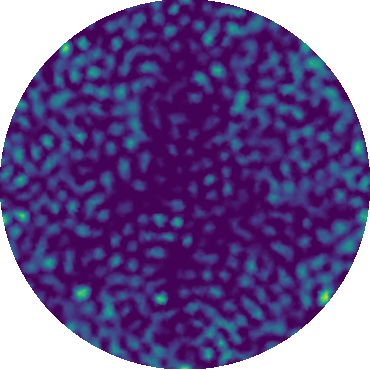}}~
    \subfloat[C][$k = 1100$]{\includegraphics[height=\scale\textwidth]{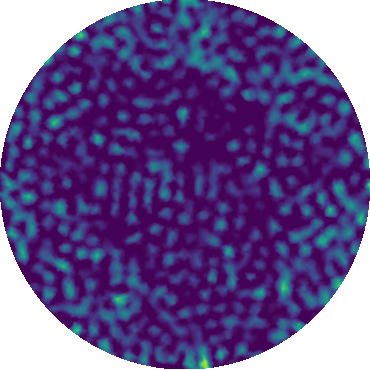}}~
    \subfloat[C][$k = 1100$]{\includegraphics[height=\scale\textwidth]{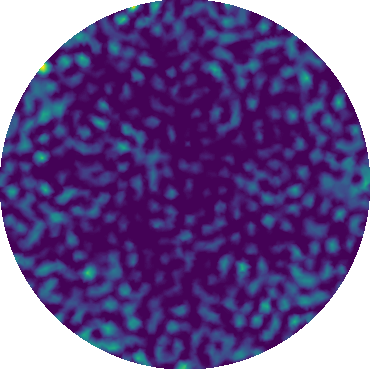}}~
    \includegraphics[height=\scale\textwidth]{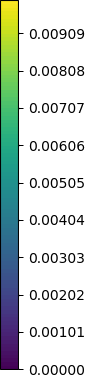}
	\caption{Reconstruction of the power densities $ H_i $ from data with an added relative noise of (row 1) 100\%, (row 2) 500\% and (row 3) 1000\%. Column $i$ correspond to boundary condition $f_i$. $k$ is the number of singular values used in the TSVD reconstruction for $H$. $ \epsilon = 0.01 $.}
    \label{fig:H0-relative-noise}
\end{figure}

In Figure \ref{fig:H0-relative-noise} we illustrate some reconstructions of the power densities $ H_i $, corresponding to the boundary conditions $ f_i $, at varying levels of relative noise; compare with Figure \ref{fig:H-true} where the true conductivities computed from the model are shown. 

Different levels of truncation of the singular values have been used in the reconstruction in an attempt to regularize and obtain the best reconstruction. These power densities are later used in the reconstruction of the conductivities. The three power density reconstructions in each row are what goes into the reconstruction of a conductivity.

The truncation level was chosen by running the full reconstructions of the conductivity for a limited range of choices and picking what seemed to be the best options. The search was not exhaustive.

We would like to draw the attention to the most significant features in Figure \ref{fig:H0-relative-noise}(a), the peaks enclosing the location of the inclusion. These are more or less defining for the ability to reconstruct the conductivity. In the whole domain remains a lot of errors background in something like the interference pattern of waves. These patterns are an artifact of the right singular vectors spanning the wave space. While they will introduce some minor errors in the conductivity reconstruction, they don't form systematic features like the peaks around the inclusion and thus shouldn't produce wrong inclusions in the final conductivity. 

We remark that the colorbars are not the same for all the power density reconstructions. The scale of the oscillations of the errors drown out more delicate features, but as we see later for the reconstructions of the conductivities some of the important parts must remain.

\subsection{Reconstructing the electrical conductivity}
We reconstruct the conductivity using the approach outlined in Section \ref{sec:Inversion} following, as mentioned there, the steps taken in \cite{jensen2018acousto} and \cite{jensen2021sound}.

\begin{figure}[!ht]
\centering
    \newcommand{\scale}{0.29}
    \subfloat[A][$k = 1500$]{\includegraphics[height=\scale\textwidth]{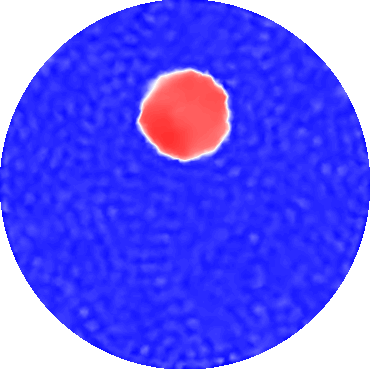}}~
    \subfloat[B][$k = 1200$]{\includegraphics[height=\scale\textwidth]{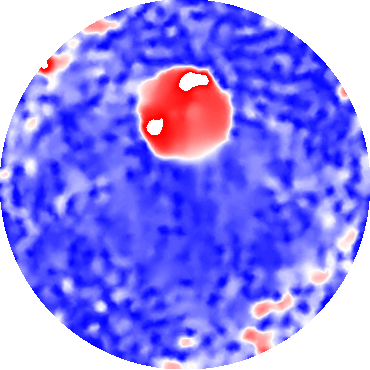}}~
    \subfloat[C][$k = 1100$]{\includegraphics[height=\scale\textwidth]{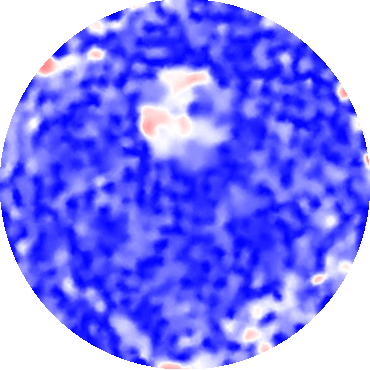}}~
    \includegraphics[height=\scale\textwidth]{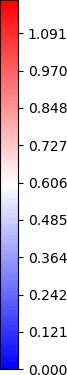}
	\caption{Reconstruction of the conductivity from data with an added relative noise of (a) 100\%, (b) 500\% and (c) 1000\%. $k$ is the number of singular values used in the TSVD reconstruction for $H$. $ \epsilon = 0.01 $.}
    \label{fig:recon-relative-noise}
\end{figure}

The reconstructions corresponding to varying levels of relative noise are shown in Figure \ref{fig:recon-relative-noise}. As mentioned in the former section, they were each reconstructed from the corresponding row-sets of power densities illustrated in Figure \ref{fig:H0-relative-noise}. We see here how robust to error the reconstruction is, with us being able to discern the inclusion even at incredibly high levels of noise.

We ascribe this largely to the level of redundancy in the data from the multiple waves used to perturb the domain. Also likely, the robustness of the model to random noise possibly exiting the model's range likely plays a part here. This latter aspect was also observed in \cite{jensen2021sound} for defects from reconstruction using an erroneous wave. There largely wrong features in the power density background were shown to have limited impact on the final reconstruction of the conductivity because they lacked a coherence across the multiple power densities going into the reconstruction.

While we do observe the reconstruction quality suffers from the heavy increase in noise levels, the inclusion remains observable even at the highest level of noise.

\section{Conclusions, discussion and further work} \label{sec:Conclusion}
In this manuscript we have suggested a mathematical and computational model for the complete inversion framework for AET. Starting with acoustic wave formation, we suggest a two step inversion procedure, first recovering  power densities as intermediate objects, then the conductivity distribution. The motivation is to understand, whether and to what extent AET can provide a modality for stable imaging of the electrical conductivity in realistic and relevant situations. Our computational phantom therefore builds from parameters motivated by medical imaging. 

We have analyzed the signal magnitude, a critical element in AET. There are many parameters affecting this quantity, including the amplitude of the pressure for ultrasound waves and electrical signals through the body. These parameters are regulated by an upper bound in clinical applications. Other being physical parameters like the coupling coefficient, which is not exactly known for many materials. Since the acoustic-electric coupling is quite weak, the signal magnitude is expected and shown to be very small. 

We have further investigated the critical and unavoidable Johnson-Nyquist noise. Due to the small amplitude signal of AET, this low amplitude background noise needs to be accounted for, regardless of the quality of the equipment and other forms of measurement noise; it can only really be dealt with by data redundancy, re-sampling and model robustness. The precise magnitude of the JN-noise is highly system dependent, but we have attempted a qualified guess on the order of magnitude. Our (uncertain) estimate of the Johnson-Nyquist noise is several orders of magnitude larger than the signal; this makes imaging tremendously difficult. 


Despite the challenging conditions, we have tested our inversion method on simulated data that are highly perturbed by noise. The results indicate that with an SNR of $0\,\si{dB}$ (relative error of $100 \%$) we achieve surprisingly good reconstructions, and even with an SNR of $-10\,\si{dB}$ (relative error of $1000 \%$), some features remain. These numerical results give us a hope that the technology may yet find new advancement to overcome current issues.

The relatively good performance in these hostile noise conditions indicates that mathematically the AET reconstruction problem is fairly well-posed, and the data contains lots of redundant information stemming from the choice of acoustic waves, the sampling of $I$, and the number of electric boundary conditions.

Our final assessment of AET is that it is unlikely to be feasible for high resolution imaging without a large improvement of the SNR, and that such an improvement can only be achieved by a very large number of repeated experiments. This is at least the case in medical imaging, where signal strength is (wisely) curbed by multiple health considerations. It might, however, be possible for AET to obtain some additional and useful information for diagnostic purpose, or for anomaly detection problems in materials where one is allowed to use higher pressure and more current. 

We leave for future studies the introduction of electrode models from EIT and a more detailed analysis of the Johnson-Nyquist noise in AET.  




\section{Acknowledgement}
Most of this work was done while BCSJ and AK were at DTU. The project was supported by the Villum Foundation
(grant no. 25893) 
\bibliographystyle{abbrv}
\bibliography{bibliography}

\end{document}